\documentclass[conference]{IEEEtran}

\usepackage{amssymb,amsmath}
\usepackage{graphics}
\usepackage{float}
\usepackage{graphicx}
\usepackage{amsmath}
\usepackage[nospace,noadjust]{cite}
\usepackage{color,soul}
\usepackage{dsfont}
\usepackage{bm}
\usepackage{upgreek}
\usepackage{arydshln}
\usepackage{enumerate}
\usepackage{graphicx}
\usepackage{threeparttable}
\usepackage{caption}
\usepackage{multirow}
\usepackage{color}
\usepackage{xfrac}
\usepackage{array}
\usepackage{nomencl}
\usepackage{hyphenat}

\newcommand{\be}{\begin{equation}}
\newcommand{\ee}{\end{equation}}
\newcommand{\innerprod}[2]{ \left\langle #1 , #2\right\rangle}
\newcommand{\parder}[2]{ \frac{\partial #1}{\partial #2} }
\newcommand{\norm}[1]{ || #1 ||}
\newcommand{\mb}[1]{\mathbf{#1}}
\newcommand{\bs}[1]{\boldsymbol{#1}}
\newcommand{\lbr}{\left\lbrace}
\newcommand{\rbr}{\right\rbrace}
\newcommand{\virg}[1]{\textquotedblleft#1\textquotedblright}

\begin{document}
%
\title{A fresh look at the Semiparametric Cram\'{e}r-Rao Bound}

\author{\IEEEauthorblockN{Stefano Fortunati\textsuperscript{1,2}, Fulvio Gini\textsuperscript{1}, Maria Greco\textsuperscript{1}}
	\IEEEauthorblockA{\textsuperscript{1}Dept. of Information Engineering\\
		University of Pisa\\
		\{s.fortunati,f.gini,m.greco\}@iet.unipi.it}
	\and
	\IEEEauthorblockN{Abdelhak M. Zoubir\textsuperscript{2}}
	\IEEEauthorblockA{\textsuperscript{2}Signal Processing Group\\
		Technische Universit\"{a}t Darmstadt\\
		zoubir@spg.tu-darmstadt.de}
	\and
	\IEEEauthorblockN{Muralidhar Rangaswamy\textsuperscript{3}}
	\IEEEauthorblockA{\textsuperscript{3}U.S. AFRL, Sensors Directorate\\
		Wright-Patterson AFB, OH, USA\\
		Muralidhar.Rangaswamy@us.af.mil}}

\maketitle

\begin{abstract}
This paper aims at providing a fresh look at semiparametric estimation theory and, in particular, at the Semiparametric Cram\'{e}r-Rao Bound (SCRB). Semiparametric models are characterized by a finite-dimensional parameter vector of interest and by an infinite-dimensional nuisance function that is often related to an unspecified functional form of the density of the noise underlying the observations. We summarize the main motivations and the intuitive concepts about semiparametric models. Then we provide a new look at the classical estimation theory based on a geometrical Hilbert space-based approach. Finally, the semiparametric version of the Cram\'{e}r-Rao Bound for the estimation of the finite-dimensional vector of the parameters of interest is provided. 
\end{abstract}

\section{Introduction}
\label{intro}
Any scientific experiment which aims to gain some knowledge about a real-word phenomenon starts with the data collection that will be later used to infer information. In statistical Signal Processing (SP) applications, all the available knowledge about the observed phenomenon is summarized in the probability distribution of the collected data. More formally, let $\mb{x}_1,\ldots,\mb{x}_M$ be $M$ measurements collected from a random experiment, then $P_0(\mb{x}_1,\ldots,\mb{x}_M)$ is their (joint) \virg{true} distribution that will be the basic ingredient of any inference method and, in particular of point estimation. In an estimation problem, in fact, we are interested in the evaluation of some functional of $P_0(\mb{x}_1,\ldots,\mb{x}_M)$, say $\nu(P_0)$. The mean value, the median, the covariance of the data are only some simple examples of $\nu(P_0)$. Clearly, any inference problem implies, as a first step, the estimation or, at least a reasonable guess, of the true data distribution $P_0(\mb{x}_1,\ldots,\mb{x}_M)$ or of the relevant probability density function (pdf) $p_0(\mb{x}_1,\ldots,\mb{x}_M)$. To this end, we have to define a family of distributions (or pdfs) that are able to statistically characterize the collected observations. The set of possible distributions for a given random experiment is called \textit{model}. According to the available amount of a-priori knowledge, two different classes of models can be exploited for data analysis and statistical inference: the \textit{parametric} and the \textit{non-parametric models}.

A parametric model $\mathcal{P}_{\bs{\theta}}$ is defined as a set of pdfs for the acquired dataset $\mb{x}_1,\ldots,\mb{x}_M$ that are parametrized by a finite-dimensional parameter vector $\bs{\theta} \in \Theta \subseteq \mathbb{R}^d$:
\be
\label{par_mod}
\mathcal{P}_{\bs{\theta}} \triangleq \lbr p_X(\mb{x}_1,\ldots,\mb{x}_M|\bs{\theta}),\bs{\theta} \in \Theta \rbr.
\ee
Clearly, if a parametric model is adopted, the knowledge about the real-word phenomenon of interest is summarized in the (finite-dimensional) parameter vector $\bs{\theta}$. In particular, if the true data pdf $p_0(\mb{x}_1,\ldots,\mb{x}_M)$ belongs to $\mathcal{P}_{\bs{\theta}}$, this implies that there exists $\bs{\theta}_0 \in \Theta$ such that $p_0(\mb{x}_1,\ldots,\mb{x}_M) = p_X(\mb{x}_1,\ldots,\mb{x}_M|\bs{\theta}_0)$ and then the model is said to be correctly specified. In a parametric model, every pdf is completely characterized by a finite number of parameters, i.e. the $d$ entries of the vector $\bs{\theta}$, and this makes the subsequent inference procedure relatively simple even when only a small number of measurements is available. However, in some practical applications, and in particular when our a-priori knowledge about the experiment at hand is limited, a parametric model could result to be too restrictive and we could run the risk that the true data pdf $p_0(\mb{x}_1,\ldots,\mb{x}_M)$ falls outside the assumed model $\mathcal{P}_{\bs{\theta}}$. If this happens, the parametric model is said to be \textit{misspecified} (\cite{SPM}, \cite{ISI2017}). In order to avoid the model misspecification (that, of course will lead to some performance degradation of the inference procedure), one can decide to characterize the statistical behavior of the collected data using a more general non-parametric model. 

A non-parametric model is a collection of pdfs that possibly satisfy some functional constraints, i.e. symmetry around their mean value, and can be indicated as:
\be
\mathcal{P}_{p} \triangleq \lbr p_X(\mb{x}_1,\ldots,\mb{x}_M) \in \mathcal{K} \rbr,
\ee 
where $\mathcal{K}$ is some constrained set of pdfs. Of course, using a non-parametric model, the risk of model misspecification is minimized since $\mathcal{P}_{p}$ is able to embrace a wider range of pdfs. On the other hand, the \virg{expanse} of $\mathcal{P}_{p}$ could represent a big issue. In fact, to obtain an estimate of the true data pdf, we have to face an infinite-dimensional estimation problem. This could represent a prohibitive task in practical applications where the number of available data is limited and it usually results to be too small to estimate the full \virg{shape} of $p_0(\mb{x}_1,\ldots,\mb{x}_M)$. 

The concept of \textit{semiparametric models} has been introduced to be a compromise between the \virg{parsimony} of parametric models and the \virg{realism} of the non-parametric ones \cite{Wellner85}. A semiparametric model is a set of pdfs characterized by a finite-dimensional parameter $\bs{\theta} \in \Theta$ along with some infinite-dimensional parameter $l \in \mathcal{L}$, where $\mathcal{L}$ is some set of functions:
\be
\label{seimp_model}
\mathcal{P}_{\bs{\theta},l} \triangleq \lbr p_X(\mb{x}_1,\ldots,\mb{x}_M|\bs{\theta},l), \bs{\theta} \in \Theta, l \in \mathcal{L}\rbr.
\ee  
In many practical applications involving semiparametric models, the parameters of interest are the ones collected in (a subvector of) the finite-dimensional vector $\bs{\theta}$, while the infinite-dimensional parameter $l$ can be considered as a \textit{nuisance} parameter, i.e. a parameter that interferes with the inference process but whose estimation is not required. Classical estimation theory has been fully developed in the case in which both the parameters of interest and the nuisance parameters can be collected in a finite dimensional vector (see e.g. \cite{kay1993fundamentalsI}, Sect. 10.7). In particular, we can easily derive a Cram\'{e}r-Rao inequality for the covariance matrix of any unbiased estimator in the presence of finite-dimensional nuisance parameters. The issue we address in this paper is the following: is it possible to generalize the classical theories in order to take an infinite-dimensional nuisance function into account?

A huge amount of practical inference problems can be described using a semiparametric formalism (see \cite{Wellner85}, \cite{BKRW}, \cite{Newey} and references therein), but its potential has not been fully exploited by the SP community as yet. To the best of our knowledge, there are only few attempts to use a semiparametric approach in SP applications. Two examples are the works \cite{Amari} and \cite{Zoubir_semi}, where a semiparametric approach has been applied to blind source separation and to nonlinear regression, respectively. Our aim here, is to provide a fresh look at semiparametric estimation and at the SCRB, that can be exploited by a wide audience of SP practitioners. 

The rest of the paper is organized as follows. In Sect. \ref{RES_model}, we show that one of the most widely used non-Gaussian model, i.e. the class of Real Elliptically Symmetric (RES) distributions (\cite{RichPhD}, \cite{Esa}), is semiparametric in nature. In Sect. \ref{par_mod_geo}, we discuss a geometrical reinterpretation of classical estimation theory and, in particular, of the CRB in the presence of a finite-dimensional nuisance vector. In Sect. \ref{Semipar}, previously introduced geometrical tools  will be used to generalize the parametric CRB to the semiparametric framework. Finally, a discussion on some open problems and on possible future research directions is provided in Sect. \ref{conclusions}.

\section{A semiparametric model: the RES distributions}
\label{RES_model}
Before providing some hints about the theory of semiparametric estimation and about the SCRB, let us discuss an example of a semiparametric model that will help us to clarify the basic ideas. We focus our attention on the class of RES distributions \cite{RichPhD}, \cite{Esa}. This family of distributions has been recognized to be one of the more suitable and general model to statistically characterize the non-Gaussian, heavy-tailed, nature of the disturbance in many practical applications \cite{Grec2}, \cite{StefJASP}.

A random vector $\mb{x} \in \mathbb{R}^N$ is said to be RES-distributed if it has a relevant pdf of the form:
\be
p_{X}(\mb{x}) = 2^{-N/2}|\bs{\Sigma}|^{-1/2}g((\mb{x}-\bs{\mu})^{T}\bs{\Sigma}^{-1}(\mb{x}-\bs{\mu})),
\ee   
where $g:\mathbb{R}_0^{+} \rightarrow \mathbb{R}^{+}$ is called \textit{density generator}. The RES class encompasses the whole family of the (real) Compound Gaussian (CG) distributions as a special case. Moreover, it contains non-Gaussian and non-CG distributions as well, as for example, the Generalized Gaussian (GG) distribution. Even if the RES class and its complex counterpart, the CES class, are two celebrated disturbance models, their intrinsic semiparametric structure has not been investigated as yet. 

In almost all practical applications involving the RES model, we are generally interested in the estimation of the mean value $\bs{\mu}$ and/or of the scatter matrix $\bs{\Sigma}$ of the RES distributions, irrespective of the particular density generator $g$. Therefore, $g$ can be considered as a nuisance function. It is immediate to verify that the RES family can be interpreted as a semiparametric model of the form:
\be
\label{CES_model}
\mathcal{P}_{\bs{\mu},\bs{\Sigma},g} \triangleq \{ p_X(\mb{x}|\bs{\mu},\bs{\Sigma},g), (\bs{\mu},\bs{\Sigma}) \in \mathbb{R}^N \times \mathcal{M}, g \in \mathcal{G}\},
\ee 
where $\mathcal{M}$ indicates the set of all the positive definite, symmetric $N \times N$ matrices and $\mathcal{G}$ indicates the set of the density generators. Due to the elliptical symmetry that characterizes the RES class, the relevant semiparametric model is a particular instance of a \textit{semiparametric group model} (see \cite{BKRW} (Sect. 4.2)). Before providing a discussion on the mathematical tools needed to handle estimation problems in the presence of an (infinite-dimensional) nuisance function, we first sketch, in the next section, a new geometrical reading of the classical estimation theory in the presence of finite-dimensional nuisance parameters that will be the basis for its generalization to the semiparametric framework.

\section{Estimation in parametric models: a geometrical interpretation}
\label{par_mod_geo}
We now introduce a geometrical interpretation of the “finite-dimensional” estimation theory that can be extended to semiparametric estimation. The classical parametric estimation theory can be reformulated using three main ingredients:
\begin{itemize}
	\itemsep0em
	\item the Hilbert space of all the zero-mean, vector-valued, random functions of the observation vectors,
	\item a notion of \textit{tangent space} for a statistical model;
	\item an orthogonal projection operator.
\end{itemize}

Before introducing these three geometrical objects, it is worth pointing out that here we assume to deal with real random vectors and real parameters. The extension to the complex field falls outside the scope of this paper. Moreover, unless otherwise stated, we assume to have a single random observation vector $\mb{x}$ at our disposal. This is not a big limitation, since the extension to the case of a set of $M$ \textit{independent} and \textit{identically distributed} (i.i.d.) observation vectors is straightforward. The generalization to the non i.i.d. case is discussed in \cite{Hallin_Werker}. Lastly, due to the lack of space, we will not provide any details about the formal definition of a Hilbert space, a linear subspace of a Hilbert space and of the projection operator. We refer the reader to the book \cite{Hilbert} or to \cite{BKRW} (Sections A.1 and A.2).

Let us start from the parametric model defined in \eqref{par_mod} where $\bs{\theta}$ is a $d$-dimensional parameter vector and $\mb{x}$ is an observation vector. We could be interested in estimating a subvector of $\bs{\theta}$. Then $\bs{\theta}$ can be partitioned as $\bs{\theta}=(\bs{\gamma}^T,\bs{\eta}^T)^T \in \Gamma \times \Omega \subseteq \mathbb{R}^q \times \mathbb{R}^r$ where the $q$-dimensional vector $\bs{\gamma}$ is the vector of the \textit{parameters of interest} while the $r$-dimensional vector $\bs{\eta}$ is the vector of the \textit{nuisance parameters}. Note that $q+r=d$. In the reminder of this section, we always indicate the \textit{true} parameter vector as $\bs{\theta}_0=(\bs{\gamma}^T_0,\bs{\eta}^T_0)^T$ and the related \textit{true} pdf as $p_0(\mb{x})\triangleq p_X(\mb{x}|\bs{\theta}_0)$. Consequently, we have that $\mb{x} \sim p_0(\mb{x})$ where $\sim$ stands for \virg{is distributed according to}. Finally, $E_0\{\cdot\}$ indicates the expectation operator taken with respect to the true pdf $p_0(\mb{x})$.

\subsection{The Hilbert space of the $q$-dimensional random functions}
\label{Hilber_random}
We introduce here a Hilbert space that plays a fundamental role in the following development. Consider the set $\mathcal{H}^q$ of all the $q$-dimensional vector-valued functions of the random vector $\mb{x}\sim p_0(\mb{x})$ such that:
\be
\label{H_space}
\mathcal{H}^q  = \{\mb{h}|\mb{h}(\mb{x}) \in \mathbb{R}^q, E_0\{\mb{h}\} = 0, E_0\{\mb{h}^T\mb{h}\}<\infty \},
\ee
where, in the expectation operator, we dropped the dependence of $\mb{h}$ on $\mb{x}$ for notation simplicity. This convention will be adopted from here onwards. It is immediate to verify that $\mathcal{H}^q$ is a Hilbert space whose inner product is defined through the expectation operator, i.e. $\innerprod{\mb{h}_1}{\mb{h}_2} = E_0\{\mb{h}_1^T\mb{h}_2\}, \forall \mb{h}_1,\mb{h}_2\in \mathcal{H}^q$ \cite{Tsiatis} (Sect. 2). Since $\mb{h}$ is a $q$-dimensional function of a random vector $\mb{x}$, we can define its $q \times q$ covariance matrix in the usual way as:
\be
\mb{C}_0(\mb{h}) \triangleq E_0\{\mb{h}\mb{h}^T\}  \in \mathbb{R}^{q \times q}.
\ee 

Let us investigate the geometrical structure of $\mathcal{H}^q$. Specifically, we focus on the derivation of an explicit expression of the orthogonal projection of a generic element $\mb{h}$ into a finite-dimensional subspace of $\mathcal{H}^q$. Following \cite{Tsiatis} (Sect. 2.4), let us define $\mb{v}=\left[ v_1, \cdots, v_k\right]^T$ as a column vector of $k$ arbitrary elements of $\mathcal{H}^1$ that is the Hilbert space of functions obtained from \eqref{H_space} by choosing $q=1$. As a finite-dimensional subspace of $\mathcal{H}^q$, we consider the linear span of vector $\mb{v}$ defined as:
\be
\label{U_subspace}
\mathcal{U} \triangleq \{\mb{A}\mb{v}:\mb{A} \mathrm{ \; is \; any \;matrix \;in  \;} \mathbb{R}^{q \times k}\}.
\ee

We want to find the orthogonal projection of an arbitrary element $\mb{h} \in \mathcal{H}^q$ onto $\mathcal{U}$, i.e. $\Pi(\mb{h}|\mathcal{U})$. From the Projection Theorem (see e.g. \cite{BKRW} (Sect. A.2)), we know that this projection exists, it is unique and it can be explicitly written as:
\be
\label{proj_lin_sub}
\Pi(\mb{h}|\mathcal{U}) = E_0\{\mb{h}\mb{v}^T\}\mb{C}_0^{-1}(\mb{v})\mb{v}.
\ee
For the sake of clarity, it is worth recalling that $\Pi(\mb{h}|\mathcal{U})$, $\mb{h}$ and $\mb{v}$ are all vector-valued functions of the random observation vector $\mb{x}$. As we will show in Subsect. \ref{CRB_geo}, this explicit expression of the projection operator is the key for a geometrical description of the Cram\'{e}r-Rao inequality in the presence of a finite-dimensional nuisance parameter vector.

\subsection{The parametric nuisance tangent space}
\label{tang_space}
In order to define the parametric nuisance tangent space, we first need to recall the notion of \textit{score vector}. Let $\mathcal{P}_{\bs{\theta}}$ be a parametric model as in \eqref{par_mod} and let $\mb{x} \sim p_0(\mb{x})$. Then, the score vector for $\mb{x}$ in $\bs{\theta}_0$, indicated as $\mb{s}_{\bs{\theta}_0} \equiv \mb{s}_{\bs{\theta}_0}(\mb{x})$, is the $d$-dimensional vector-valued function whose entries, for $i=1,\ldots,d$, are defined as:
\be
\label{score_vect_ver1}
\left[ \mb{s}_{\bs{\theta}_0} \right]_i \triangleq  \left[ \nabla_{\bs{\theta}}\ln p_X(\mb{x}|\bs{\theta}_0)\right]_i= \left. \parder{\ln p_X(\mb{x}|\bs{\theta})}{\theta_i} \right|_{\bs{\theta}=\bs{\theta}_0}. 
\ee 
Note that, if $\bs{\theta}_0=(\bs{\gamma}_0^T,\bs{\eta}_0^T)^T$, the score vector in \eqref{score_vect_ver1} can be partitioned as $\mb{s}_{\bs{\theta}_0} = \left(\mb{s}_{\bs{\gamma}_0}^T, \mb{s}_{\bs{\eta}_0}^T \right)^T$.
Under the usual regularity conditions that allow for the order inversion between integral and derivative operators, the score vector is a zero-mean random vector, i.e. $E_0\{\mb{s}_{\bs{\theta}_0}\}=\mb{0}$ and each of its entries has finite variance, i.e. $E_0\{[\mb{s}_{\bs{\theta}_0}]_i^2\}<\infty$. Consequently, each entry of $\mb{s}_{\bs{\theta}_0}$ belongs to $\mathcal{H}^1$ obtained from \eqref{H_space} with $q=1$. Then, using the procedure discussed in Subsection \ref{Hilber_random}, we can define a finite-dimensional subspace of $\mathcal{H}^q$ (see \eqref{U_subspace}) as the linear span generated by the entries of the nuisance score vector $\mb{s}_{\bs{\eta}_0}$. In particular, we define the \textit{nuisance tangent space} as the finite-dimensional subspace of $\mathcal{H}^q$ spanned by the entries of $\mb{s}_{\bs{\eta}_0}$, i.e.
\be
\label{tangent_space_nuisance}
\mathcal{T}_{\bs{\eta}_0} \triangleq \{\mb{C}\mb{s}_{\bs{\eta}_0}:\mb{C} \mathrm{ \; is \; any \;matrix \;in  \;} \mathbb{R}^{q \times r}\},
\ee
where $q=\dim(\bs{\gamma})$ and $r=\dim(\bs{\eta})$. 

\subsection{The efficient score vector and the Cram\'{e}r-Rao Bound}
\label{CRB_geo}
As before, let $\mathcal{P}_{\bs{\theta}}$ be a parametric model as in \eqref{par_mod} and let $\mb{x} \sim p_0(\mb{x})$ be the random observation vector, then the Fisher Information Matrix (FIM) for $\bs{\theta}_0$ is defined as follows:
\be
\label{FIM}
\mb{I}(\bs{\theta}_0) = E_0\{\mb{s}_{\bs{\theta}_0}\mb{s}^T_{\bs{\theta}_0}\} = \left( \begin{array}{cc}
	\mb{C}_0(\mb{s}_{\bs{\gamma}_0}) & \mb{I}_{{\bs{\gamma}_0}{\bs{\eta}_0}}\\
	\mb{I}_{{\bs{\gamma}_0}{\bs{\eta}_0}}^T   &  \mb{C}_0(\mb{s}_{\bs{\eta}_0})
\end{array}\right) 
\ee
where $\mb{I}_{{\bs{\gamma}_0}{\bs{\eta}_0}} \triangleq E_{p_0}\{\mb{s}_{\bs{\gamma}_0}\mb{s}_{\bs{\eta}_0}^T\}$ (see e.g. \cite{kay1993fundamentalsI}). Moreover, let $\hat{\bs{\gamma}}(\mb{x})$ be an \textit{unbiased} estimator of the vector of the parameter of interest $\bs{\gamma}_0$ in the presence of the nuisance vector $\bs{\eta}_0$. Then, the Cram\'{e}r-Rao inequality on the error covariance matrix of $\hat{\bs{\gamma}}(\mb{x})$ can be easily established as:
\be
\label{CRB_n}
\begin{split}
	E_{p_0}&\{(\hat{\bs{\gamma}}(\mb{x})-\bs{\gamma}_0)(\hat{\bs{\gamma}}(\mb{x})-\bs{\gamma}_0)^T\} \geq \mathrm{CRB}(\bs{\gamma}_0|\bs{\eta}_0) \\
	&\triangleq \left(\mb{C}_{p_0}(\mb{s}_{\bs{\gamma}_0}) - \mb{I}_{{\bs{\gamma}_0}{\bs{\eta}_0}} \mb{C}_0^{-1}(\mb{s}_{\bs{\eta}_0})\mb{I}_{{\bs{\gamma}_0}{\bs{\eta}_0}}^T \right)^{-1}.
\end{split}
\ee
In particular, $\mathrm{CRB}(\bs{\gamma}_0|\bs{\eta}_0)$ can be obtained as the top-left submatrix of the inverse of the FIM $\mb{I}(\bs{\theta}_0)$ in \eqref{FIM} whose explicit expression, given in \eqref{CRB_n}, follows directly from the application of the Matrix Inversion Lemma. 
Interestingly enough, this result can also be obtained using the geometrical approach discussed in Subsects. \ref{Hilber_random} and \ref{tang_space}. To prove this, we firstly introduce the notion of \textit{efficient score vector} (\cite{BKRW} (Sect. 2) and \cite{Tsiatis} (Sect. 3.4)).
The efficient score vector $\mb{s}_0^{\star} \equiv \mb{s}_0^{\star}(\mb{x})$ is defined as the residual of the score vector of the parameters of interest $\mb{s}_{\bs{\gamma}_0}$ after projecting it onto the nuisance tangent space $\mathcal{T}_{\bs{\eta}_0}$ defined in \eqref{tangent_space_nuisance}:
\be
\label{efficient_score_vect}
\begin{split}
	\mb{s}_0^{\star} & \triangleq \mb{s}_{\bs{\gamma}_0} - \Pi(\mb{s}_{\bs{\gamma}_0}|\mathcal{T}_{\bs{\eta}_0})\\
	& = \mb{s}_{\bs{\gamma}_0} - E_{p_0}\{\mb{s}_{\bs{\gamma}_0}\mb{s}^T_{\bs{\eta}_0}\}\mb{C}_0^{-1}(\bs{\eta}_0)\mb{s}_{\bs{\eta}_0},
\end{split}
\ee
where, for the last equality, we used the explicit projection formula derived in \eqref{proj_lin_sub}. Roughly speaking, the efficient score vector can be used to quantify the amount of information carried by the true data pdf, $p_X(\mb{x}|\bs{\theta}_0)$, about the vector of parameters of interest $\bs{\gamma}_0$ when the nuisance vector $\bs{\eta}_0$ is unknown. This information can be summarized in the \textit{efficient FIM} defined as:
\be
\label{Eff_FIM}
\begin{split}
	\mb{I}^{\star}(\bs{\theta}_0) & = \mb{C}_0(\mb{s}_0^{\star}) = E_0\{\mb{s}_0^{\star}(\mb{s}_0^{\star})^T\} \\
	& = \mb{C}_0(\mb{s}_{\bs{\gamma}_0}) - \mb{I}_{{\bs{\gamma}_0}{\bs{\eta}_0}} \mb{C}_0^{-1}(\mb{s}_{\bs{\eta}_0})\mb{I}_{{\bs{\gamma}_0}{\bs{\eta}_0}}^T, 
\end{split}
\ee
where the last equality follows by substituting the expression of the efficient score vector, given in \eqref{efficient_score_vect}, in the expectation operator in \eqref{Eff_FIM}. It is immediate to verify that the inverse of the efficient FIM is equal to the CRB on the estimation of $\bs{\gamma}_0$ in the presence of the nuisance vector $\bs{\eta}_0$ given in \eqref{CRB_n}, i.e.
\be
\label{CRB_eff_FIM}
\mathrm{CRB}(\bs{\gamma}_0|\bs{\eta}_0) = (\mb{I}^{\star}(\bs{\theta}_0))^{-1}.
\ee
Equality in \eqref{CRB_eff_FIM} provides the link between the classical approach to estimation theory and the geometrical one, introduced in Subsects. \ref{Hilber_random} and \ref{tang_space}. More important, it shows us that, in order to derive a CRB for estimation problems in the presence of nuisance parameters, we only need two geometrical objects: the nuisance tangent space $\mathcal{T}_{\bs{\eta}_0}$ and a projection operator on $\mathcal{T}_{\bs{\eta}_0}$, i.e. $\Pi(\cdot|\mathcal{T}_{\bs{\eta}_0})$. Remarkably, none of the previous geometrical objects requires the finite-dimensionality of the nuisance parameters, hence they can be readily applied to the semiparametric framework. 

\section{Extension to Semiparametric Models}
\label{Semipar}
The geometrical framework previously introduced can be extended to the semiparametric case. In order to maintain the notation as consistent as possible with the one used in Sect. \ref{par_mod_geo}, we define as \textit{semiparametric model} the set of densities:
\be
\label{def_semipar_model}
\mathcal{P}_{\bs{\gamma},l} \triangleq \lbr p_X(\mb{x}|\bs{\gamma},l),\bs{\gamma} \in \Gamma \subseteq \mathbb{R}^q, l \in \mathcal{L} \rbr,
\ee
where, as before, $\bs{\gamma}$ is a $q$-dimensional vector of the parameters of interest, while $l$ is a nuisance function belonging to some set $\mathcal{L}$. We denote the true \virg{semiparametric vector} as $(\bs{\gamma}_0^T,l_0)^T \in \Gamma \times \mathcal{L}$, and consequently the true pdf is $p_0(\mb{x}) \triangleq p_X(\mb{x}|\bs{\gamma}_0,l_0)$. Clearly, if we set the unknown nuisance function $l \in \mathcal{L}$ to be the true one, i.e. $l_0$, the model $\mathcal{P}_{\bs{\gamma},l_0} \equiv \mathcal{P}_{\bs{\gamma}}$ can be considered as a parametric model, where each pdf is indexed by the finite-dimensional parameter vector $\bs{\gamma} \in \Gamma$. As discussed in Sect. \ref{intro}, we cannot directly apply inference methods developed in the parametric framework to the semiparametric one because of the non-parametric, infinite-dimensional nature of the nuisance function $l \in \mathcal{L}$. This dimensionality problem can be overcome by introducing the concept of \textit{parametric submodel} of a semiparametric model (\cite{BKRW} (Sect. 3.1, Def. 1) and \cite{Tsiatis} (Sect. 4.2)). Specifically, the \textit{i-th} parametric submodel of the semiparametric model $\mathcal{P}_{\bs{\gamma},l}$, denoted as:
\be
\mathcal{P}_{\bs{\gamma},\nu_i(\bs{\eta})}  = \lbr p_X(\mb{x}|\bs{\gamma},\nu_i(\bs{\eta})),\bs{\gamma} \in \Gamma, \bs{\eta} \in \Omega_i \rbr,
\ee
is defined as a class of parametric pdfs indexed by a finite-dimensional parameter vector $(\bs{\gamma}^T,\bs{\eta}^T)^T \in \Gamma \times \Omega_i \subseteq \mathbb{R}^q \times \mathbb{R}^{r_i}$, such that, for every $i \in I$:
\begin{itemize}
	\item[C0)] $\nu_i(\bs{\eta}): \Omega_i \rightarrow \mathcal{L}$ is a smooth parametric map,
	\item[C1)] $\mathcal{P}_{\bs{\gamma},\nu_i(\bs{\eta})} \subseteq \mathcal{P}_{\bs{\gamma},l}$,
	\item[C2)] $p_0(\mb{x}) \in \mathcal{P}_{\bs{\gamma},\nu_i(\bs{\eta})}$, i.e. there exists a vector $(\bs{\gamma}^T_0,\bs{\eta}^T_{0})^T$ such that $p_X(\mb{x}|\bs{\gamma}_0,\nu_i(\bs{\eta}_0)) = p_X(\mb{x}|\bs{\gamma}_0,l_0) \triangleq p_0(\mb{x})$.
\end{itemize}
Condition C1 tells us that all the pdfs that compose each possible parametric submodel $\mathcal{P}_{\bs{\gamma},\nu_i(\bs{\eta})}$ have to belong to the semiparametric model $\mathcal{P}_{\bs{\gamma},l}$ as well. Moreover, Condition C2 highlights the fact that each parametric submodel $\mathcal{P}_{\bs{\gamma},\nu_i(\bs{\eta})}$ must contain the true pdf $p_0(\mb{x})$. Roughly speaking, by using a parametric submodel $\mathcal{P}_{\bs{\gamma},\nu_i(\bs{\eta})}$ in place of $\mathcal{P}_{\bs{\gamma},l}$ we are actually identifying the infinite-dimensional parameter $l \in \mathcal{L}$ with the finite-dimensional nuisance parameter vector $\bs{\eta} \in \Omega_i \subseteq \mathbb{R}^{r_i}$ whose dimension $r_i$ depends on the choice of the particular parametric submodel. The way to generalize the theory developed for parametric models to semiparametric models should now be clear. The idea is to exploit the finite-dimensional statistical results in the (artificial) set of parametric submodels $\{\mathcal{P}_{\bs{\gamma},\nu_i(\bs{\eta})}\}_{i \in I}$ and then \virg{take the limit} to generalize them to the infinite-dimensional semiparametric framework. 

\subsection{The semiparametric nuisance tangent space}
\label{semi_tan_space}
Let us now define a key element of the semiparametric theory, i.e. the semiparametric nuisance tangent space $\mathcal{T}_{l_0}$, according to the definition given in \cite{Newey} and \cite{Tsiatis} (Sect. 4.4). Note that a more general (but more abstract) definition is given in \cite{BKRW} (Sect. 3.2) and in \cite{Begun}.
At first, let us recall that the Hilbert space 
$\mathcal{H}^q$ in \eqref{H_space} is a metric space with squared distance given by $\norm{\mb{h}_1-\mb{h}_2}^2 = E_0\{(\mb{h}_1-\mb{h}_2)^T(\mb{h}_1-\mb{h}_2)\}$. The semiparametric nuisance tangent space $\mathcal{T}_{l_0}$ of the semiparametric model $\mathcal{P}_{\bs{\gamma},l}$ is defined as the closure of the union of all nuisance tangent spaces 
\be
\label{nuisance_tang_space_sub}
\mathcal{T}_{\bs{\eta}_{0,i}} \triangleq \{\mb{C}_i\mb{s}_{\bs{\eta}_{0,i}}:\mb{C}_i \mathrm{ \; is \; any \;matrix \;in  \;} \mathbb{R}^{q \times r_i}\}
\ee
of the parametric submodels $\{\mathcal{P}_{\bs{\gamma},\nu_i(\bs{\eta})}\}_{i \in I} \subseteq \mathcal{P}_{\bs{\gamma},l}$. Specifically, $\mathcal{T}_{l_0} \subseteq \mathcal{H}^q$ is the subspace of all $q$-dimensional, zero-mean, vector-valued, random functions $\mb{t} \in \mathcal{H}^q$ for which there exists a sequence $\{\mb{C}_i\mb{s}_{\bs{\eta}_{0,i}}\}_{i \in I} $ such that:
\be
\label{limit_points}
\norm{\mb{t}-\mb{C}_i\mb{s}_{\bs{\eta}_{0,i}}} \underset{i\in I}{\longrightarrow} 0,
\ee
where $\mb{s}_{\bs{\eta}_{0,i}}$ is the nuisance score vector of the parametric submodel $\mathcal{P}_{\bs{\gamma},\nu_i(\bs{\eta})}$ and the matrices $\mb{C}_i$ have appropriate dimensions, i.e. $\mb{C}_i \in \mathbb{R}^{q \times r_i}$ when the nuisance parameter vector $\bs{\eta}$ belongs to $\mathbb{R}^{r_i}$. Using this definition, the semiparametric nuisance tangent space can be simply set up as:
\be
\label{semi_tangent_space_def}
\mathcal{T}_{l_0} = \overline{ \underset{\{\mathcal{P}_{\bs{\gamma},\nu_i(\bs{\eta})}\}_{i \in I}  }{\bigcup}\mathcal{T}_{\bs{\eta}_{0,i}}}.
\ee
Note that the closure $\overline{\mathcal{A}}$ of a set $\mathcal{A}$ is defined as the smallest closed set that contains
$\mathcal{A}$, or equivalently, as the set of all elements in $\mathcal{A}$ together with all the limit
points of $\mathcal{A}$. In our case, the limit points are those defined by the convergence points of all the sequences in \eqref{limit_points}. The semiparametric nuisance tangent space $\mathcal{T}_{l_0}$ is assumed to be a closed and linear subspace of the Hilbert space $\mathcal{H}^q$ (see \cite{Begun} (Assumption S), \cite{Tsiatis} (Sect. 4.4, Remark 5)). Then, the existence and the uniqueness of the orthogonal projection operator onto $\mathcal{T}_{l_0}$, i.e. $\Pi(\cdot|\mathcal{T}_{l_0})$, is guaranteed by the Projection Theorem.

\subsection{The Semiparametric Cram\'{e}r-Rao Bound}

The projection operator $\Pi(\cdot|\mathcal{T}_{l_0})$ leads us to the definition of the semiparametric counterpart of the CRB, reported in Theorem 1.

\textbf{Theorem 1}:
The Semiparametric Cram\'{e}r-Rao Bound (SCRB) for the estimation of the finite-dimensional vector $\bs{\gamma}_0$ in the presence of the nuisance function $l_0\in \mathcal{L}$ is given by:
\be
\label{SEB}
\mathrm{SCRB}(\bs{\gamma}_0|l_0)\triangleq \sup_{\{\mathcal{P}_{\bs{\gamma},\bs{\eta}_i}\}_{i \in I}} \mb{C}_0^{-1}(\mb{s}_{0,i}^{\star}) = [\bar{\mb{I}}(\bs{\gamma}_0|l_0)]^{-1},
\ee
where $\mb{s}_{0,i}^{\star}=\mb{s}_{\bs{\gamma}_0} - \Pi(\mb{s}_{\bs{\gamma}_0}|\mathcal{T}_{\bs{\eta}_{0,i}})$ (see \eqref{efficient_score_vect}) is the efficient score vector for the \textit{i-th} parametric submodel $\mathcal{P}_{\bs{\gamma},\bs{\eta}_i}$ while 
\be
\label{S_E_FIM}
\bar{\mb{I}}(\bs{\gamma}_0|l_0) \triangleq E_0\{\bar{\mb{s}}_0(\bar{\mb{s}}_0)^T\},
\ee
is the \textit{semiparametric efficient FIM} and $\bar{\mb{s}}_0\equiv\bar{\mb{s}}_0(\mb{x})$ is the \textit{semiparametric efficient score vector} defined as:
\be
\label{Semi_eff_score}
\bar{\mb{s}}_0 = \mb{s}_{\bs{\gamma}_0} - \Pi(\mb{s}_{\bs{\gamma}_0}|\mathcal{T}_{l_0}),
\ee
where $\Pi(\mb{s}_{\bs{\gamma}_0}|\mathcal{T}_{l_0})$ is the orthogonal projection of $\mb{s}_{\bs{\gamma}_0}$ on the semiparametric nuisance tangent space. Note that $\bar{\mb{s}}_0$, $\mb{s}_{\bs{\gamma}_0}$ and $\Pi(\mb{s}_{\bs{\gamma}_0}|\mathcal{T}_{l_0})$ are, in general, $q$-dimensional vector-valued random functions of the observation vector $\mb{x}$.

This theorem can be found in \cite{Tsiatis} (Theorem 4.1) and in \cite{Newey}. A more abstract and general formulation can be found in \cite{Begun} and in \cite{BKRW} (Section 3.4).

Two comments are in order. First of all, it is immediate to verify that the expression of the SCRB in \eqref{SEB} and of the semiparametric efficient score function in \eqref{Semi_eff_score} are formally equivalent to the ones introduced in Sect. \ref{CRB_geo} in the case of finite-dimensional nuisance parameters. The only difference is in the definition of the nuisance tangent space. This confirms the intuition that, from and abstract and geometrical standpoint, the parametric and the semiparametric frameworks are equivalent. Secondly, from \eqref{SEB}, it is clear that $\mathrm{SCRB}(\bs{\gamma}_0|l_0)$ is higher that any $\mathrm{CRB}(\bs{\gamma}_0|\bs{\eta}_{0,i})=\mb{C}_0^{-1}(\mb{s}_{0,i}^{\star})$ derived for the \textit{i-th} parametric submodel. In words, this means that a semiparametric model contains less information on the parameter vector of interest $\bs{\gamma}_0$ than any of its possible parametric submodel. 

The SCRB is of great practical usefulness since it provides a lower bound to the error covariance of any \textit{robust} estimator of the finite-dimensional parameter vector $\bs{\gamma}_0$, i.e. of any estimator of $\bs{\gamma}_0$ that does not rely on the a-priori knowledge of the nuisance function $l_0$. For example, the SCRB in \eqref{SEB} can be used to obtain a lower bound for \textit{any} robust estimator of the scatter matrix $\bs{\Sigma}$ of a set of RES-distributed observation vectors  (\cite{Hallin_P_2006}, \cite{PAINDAVEINE}).

\section{Conclusion}
\label{conclusions}
The aim of this paper is to provide a fresh look at semiparametric estimation, and in particular at the Semiparametric Cram\'{e}r-Rao Bound (SCRB), that can be usable by a wide audience of SP practitioners. In particular, the possible applications and the potential advantages, that we may have in addressing classical problems by using this approach, have been discussed. Of course, a huge amount of work still remain to be done. Among the numerous open issues, the biggest challenge for the application of the semiparametric theory to practical inference problems is the calculation of the projection operator $\Pi(\cdot|\mathcal{T}_{l_0})$. The monograph \cite{BKRW} provided us with many examples of this calculation for different semiparametric models. A different and more general (since it can be applied also in non i.i.d. cases) approach is discussed in \cite{Hallin_Werker}. Our current effort is devoted to the investigation of semiparametric inference methods in the context of RES and CES distributions. In particular, we are looking for a closed form expression of the SCRB for the estimation of the mean vector and the scatter matrix to assess the performance of various robust $M$-estimators. 


\section*{Acknowledgment}
The work of Stefano Fortunati has been supported by the Air Force Office of Scientific Research under award number FA9550-17-1-0065.

\bibliographystyle{IEEEtran}
\bibliography{ref_EUSIPCO}

\end{document}